%% file: asop-paper.tex
\documentclass[10pt,conference,a4paper]{IEEEtran}
\usepackage{cite}
\usepackage{graphicx}
\usepackage{xcolor} 
\usepackage{svg} 
\usepackage{url}
\usepackage[linesnumbered, ruled]{algorithm2e} 
\usepackage{amsmath}
\usepackage{bm} 
\usepackage{mathrsfs} 
\usepackage{enumerate} 
\usepackage[T1]{fontenc}

\SetCommentSty{mycommfont}
\IEEEoverridecommandlockouts

\graphicspath{ {images/} }

\begin{document}


\title{ASOP: A Sovereign and  Secure Device Onboarding Protocol for  Cloud-based IoT Services }


\author{
\IEEEauthorblockN{Khan Reaz, Gerhard Wunder}
\IEEEauthorblockA{Cybersecurity and AI Research Group\\
    Freie Universit\"at Berlin, Germany \\
    Email: khanreaz@ieee.org, g.wunder@fu-berlin.de
    }
    \thanks{This work is carried out within \textit{``PHY2APP: Erweiterung von Physical Layer Security für Ende-zu-Ende Absicherung des IoT"} project, which is funded by the German Federal Ministry of Education and Research (BMBF) under grant number 16KIS1473~\cite{phy2app}.}
}
        
\maketitle


\begin{abstract}
\input{sections/0_abstract}
\end{abstract}

\begin {IEEEkeywords}
Application-layer onboarding, Device onboarding, FIDO Specification, CRYSTALS, Post Quantum Cryptography
\end{IEEEkeywords}

\input{sections/1_introduction.tex}
\input{sections/2_section.tex}
\input{sections/3_section.tex}
\input{sections/5_conclusion.tex}

\bibliographystyle{IEEEtran}
\bibliography{asop-paper}

\end{document}

%% file: sections/0_abstract.tex
The existing high-friction device onboarding process hinders the promise and potentiality of Internet of Things (IoT). Even after several attempts by various device manufacturers and working groups, no widely adopted standard solution  came to fruition. The latest attempt by Fast Identity Online (FIDO)  Alliance~\cite{fido2021} promises a \textit{`zero touch'} solution for mass market IoT customers, but the burden is transferred to the intermediary supply chain (i.e. they have to maintain infrastructure for managing keys and digital signatures called  \textit{`Ownership Voucher'} for all devices). The specification relies on a \textit{`Rendezvous Server'}  mimicking the notion of \textit{`Domain Name System (DNS) server'}. This essentially means resurrecting all existing possible attack scenarios associated with DNS, which include Denial of Service (DoS) attack, and Correlation attack. \textit{`Ownership Voucher'} poses the risk that  some intermediary supply chain agents may act maliciously and reject the transfer of ownership or sign with a wrong key. Furthermore, the deliberate use of the weak elliptic curve \texttt{SECP256r1/SECP384r1} (also known as NIST P-256/384)~\cite{safecurves2013} in the specification raises questions. As part of the ongoing work under the German Federal Ministry of Education and Research funded project PHY2APP~\cite{phy2app},  we introduce \textsc{ASOP:} a sovereign and secure device onboarding  protocol for IoT devices without blindly trusting the device manufacturer, supply chain, and cloud service provider. The ASOP protocol allows onboarding an IoT device to a cloud server with the help of an authenticator owned by the user. This paper outlines the preliminary development of the protocol and its high-level description.  Our \textit{`zero-trust'} and \textit{`human-in-the-loop'} approach guarantees that the device owner does not remain at the mercy of  third-party infrastructures, and it utilises recently standardized post-quantum cryptographic suite (CRYSTALS) to secure connection and messages.


%% file: sections/1_introduction.tex
\section{Introduction}
\label{intro}

Internet of Things (IoT)  deployment involves the installation of the physical device and the setup of credentials  so that it can securely communicate with its target cloud or platform. This high-friction on-boarding process hinders the adoption of IoT devices-- even though a plethora of studies forecast that worldwide spending on IoT technologies will reach \$1.2 trillion in 2022~\cite{IDC2018}. Although multiple companies and  working groups have tried to automate the on-boarding process~\cite{fido2021},~\cite{intelsdo},~\cite{oath2reference}, until now there has not been a widely accepted industry standard. NIST  started a project in 2020 aiming to mitigate this well-known challenge~\cite{nistnetwork-onboarding}. Many solutions that do exist require that the end customer  be known at the time of the device manufacture (specifically during the silicon fabrication time) so that the device can be pre-configured. On top of that, they may require a discrete secure element to be mounted on those  devices for safe credentials storage. The \textit{Achilles' heel} of this process lies in the silicon fabrication time when the device root-key is  generated. After the scandalous reports on some leading 5G equipment manufacturers-- the world's view has shifted from blindly trusting device manufacturers since they can be intimidated by the state to hand over the copy of the root-keys or to install a backdoor to espionage foreign entities or nationals. Moreover, the current Public Key Infrastructure-based  models in practice prohibit users from having complete control over their device's identity data. Users’ device identity attributes are stored on multiple centralised infrastructures and multi-layer PKIs. There is a need for \textit{ `zero-trust'} and \textit{`human-in-the-loop'} solution to ease the complexity of device  onboarding for the consumer mass market.

In the German Federal Ministry of Education and Research funded PHY2APP project~\cite{phy2app}, we are developing an end-to-end device onboarding solution leveraging    Physical Layer Security and classical cryptography. Here, an off-the-shelf (OTS) IoT device is connected to a cloud-based back-end analytics provider without being tied to the device manufacturer's Public Key Infrastructure (PKI) or their Software Development Kit (SDK). Most IoT devices require two different onboarding stages: one at the network layer, which enables them to connect securely to the network; and one at the application layer, which enables them to become operational at the application layer. In~\cite{reazComPassIMIS2021}, we described \textit{ComPass} protocol to achieve network-layer onboarding by utilising the reciprocity property of a wireless channel. In  this paper, we provide a high-level description of the ongoing development of an application-layer onboarding protocol. The term, onboarding, when used alone in this paper,  refers to application-layer onboarding.


%% file: sections/2_section.tex
\section{Review of FIDO Device Onboard Specification}
\label{section2}

The FIDO Alliance was established in 2013 with an aim to make the world less dependent on passwords to  use  websites~\cite{fido2013}. Since then they have published several specifications  and protocol suites allowing biometrics and hardware token based authentication. 

\begin{figure}[!ht]
    \centering
    \includegraphics[scale=0.45]{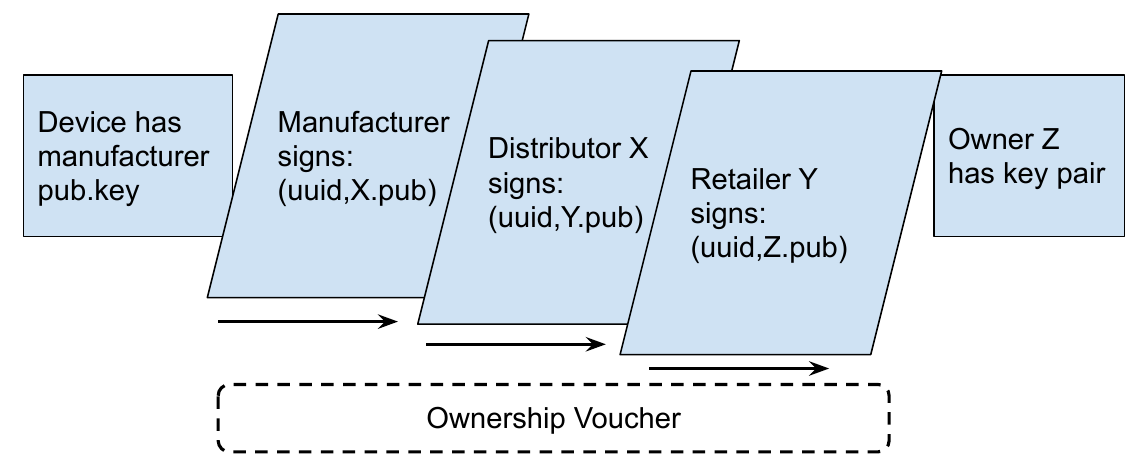}
    \caption{FIDO Ownership Voucher handover flow}
    \label{fig:FIDO_voucher}
\end{figure}

\begin{figure}[!ht]
    \centering
    \includegraphics[scale=0.36]{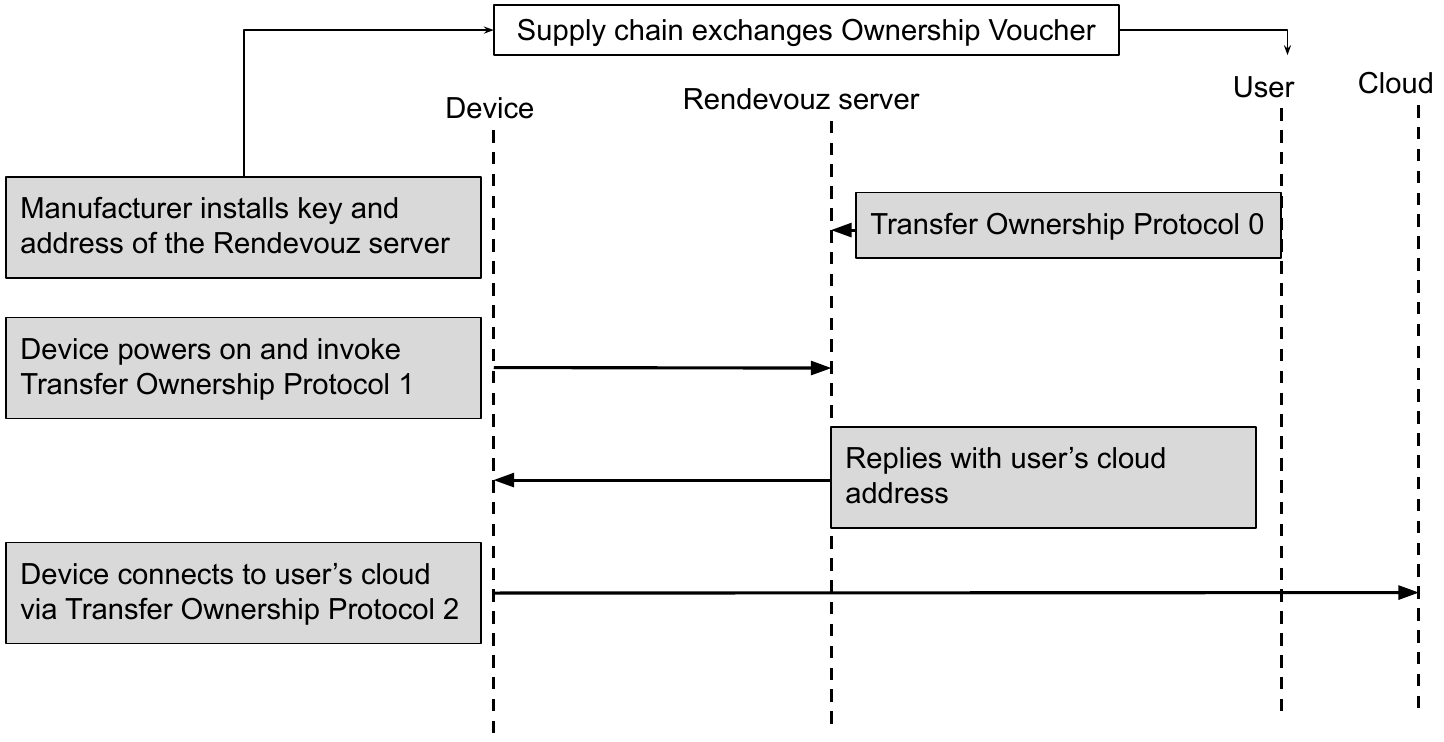}
    \caption{Graphical representation of the FIDO Device Onboard Protocols}
    \label{fig:FIDO_protocol}
\end{figure}

In recent years, they attempted to create another standard for IoT devices: FIDO Device Onboard (FDO)~\cite{fido2021}. At its core, FDO has two crucial components: (1) Ownership Voucher, and (2) Rendevouz Server. An Ownership Voucher is a digital chained certificate originated by the manufacturer where the device's unique ID and the next distributor's public key are  signed by the manufacturer. Next, when the distributor ships the device to a retailer, it is signed with the retailer's public key. This process is done for all intermediary supply chain partners. In the last stage, the seller needs to know the public key of the prospective user to sign it. Notice that every party needs to know the public key of the next party a priori as depicted in Fig.~\ref{fig:FIDO_voucher}. When the user buys a device, s/he also receives the Ownership Voucher and the seller sends the copy of it to the Rendevouz Server. The Rendevouz server is the first entry point for a device after the network-onboarding phase. Once a device is powered up, it queries the Rendevouz Server to find its user's intended IoT analytics provider's server address (intermediate steps are shown in Fig.~\ref{fig:FIDO_protocol}). In FDO, \texttt{SECP256r1/SECP384r1} elliptic curve is used even though these curves have known weaknesses as found out by Berstein et al. in~\cite{safecurves2013}. The avoidance of  well-known superior elliptic curves such as \textit{Curve25519} raises questions about the intention of the FIDO Alliance.

%% file: sections/3_section.tex
\section{ Protocol Description }
\label{section3}

\subsection{Notation}

We describe an entity $X$ which generates a public key pair to be used for another entity $Y$ as $ (X^{Y}_{s}, X^{Y}_{p}) $ where $ X^{Y}_{s}$ is the secret key and  $ X^{Y}_{p} $ is the public key. Message $(M)$ from source $ X $ to destination $ Y $ flows from left to right and the encrypted message  concatenated with  $z$ is written as $ X \rightarrow Y: Y^{X}_{p}(M) || z $.
\subsection{Entities}

Throughout the document, it is  assumed that the devices use  internet protocol (IP) to communicate with each other. In case of non-IP access methods such as Bluetooth low energy or Zigbee, the device connects via IP gateway.

\subsubsection{User $(U)$}The user, or often known as end-user, is someone who actually  uses  the device after purchasing it from some reseller. For typical smart homes and businesses, the user is a human, capable of operating electronic equipment such as a smartphone. In this paper, we do not consider the industrial use case where  IoT devices are used at large scale; in that case a user might not be a human entity.

\subsubsection{Device $(D)$} A device is being manufactured in various factories. These factories are often perceived as Original Equipment Manufacturer (OEM) who produce parts and assemble them on behalf  of other companies. During the fabrication  of the silicon chip, the OEM has to create a root-of-trust identity for that particular device; this is a universally unique identifier (UUID). Depending on the device type,  a root certificate  may also be generated using public key cryptography that is chained with the OEM's own certificate(s).

\subsubsection{Authenticator $(A)$} An authenticator is a handheld smart device such as a smartphone or a tablet. It has a rich user interface and it can communicate via different radio interfaces such as Wi-Fi, cellular network, Bluetooth etc.  The user primarily operates the authenticator on his/her own, meaning the device needs to be authenticated by some password, pattern or facial recognition techniques before usage.

\subsubsection{Server $(S)$}An IoT analytics provider operates their services  in their own cloud or in a hosted cloud architecture. We refer to the physical or virtual instance of that service as a server. The server is available via internet or in case of local installation, it can be discovered within the local network. Server API's are protected behind authentication and authorisation.

We assume that the network-level onboarding is already performed using \textit{ComPass} protocol~\cite{reazComPassIMIS2021}. Hence, the authenticator and the device can securely exchange messages using the mutually generated symmetric key $(C_K)$.

\subsection{Entity Interactions}

\begin{figure}[htbp]
    \centering
    \includegraphics[scale=0.38]{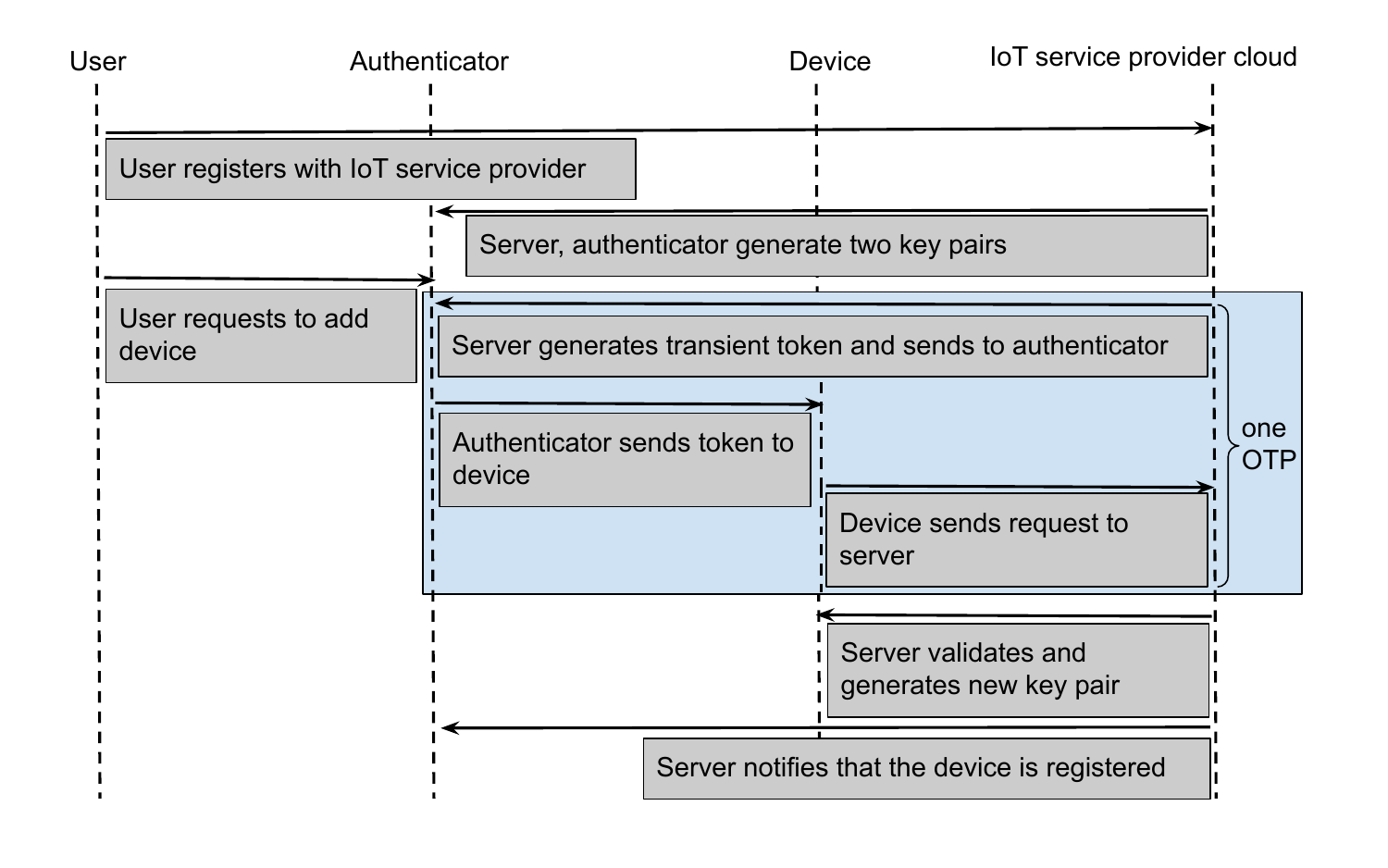}
    \caption{High level message flow diagram of ASOP protocol }
    \label{fig:protocol_flow}
\end{figure}

    For example, a user buys some mass produced IoT device from the market. It could be a smart IP camera to secure his/her house from burglary and to get better incentive from the home insurance provider. The user wishes to use an AI powered surveillance  service that is offered by some IoT analytics provider running on a cloud server. The user chooses a company and downloads their app on a smartphone or tablet. The app and the smartphone together serve as the authenticator. The user registers with the IoT analytics provider from the authenticator. This allows the user to access the server API to add/update devices and use other services provided by the IoT analytics company.
    During the account registration phase, the server and authenticator generate and exchange ephemeral root key pairs: $(S^{A}_{s}, S^{A}_{p})$,$(A^{S}_{s}, A^{S}_{p})$ using the newly standardised quantum attack resistant public key algorithm CRYSTALS-KYBER~\cite{crystals}. These keys have a certain validity period, ideally in the range of hours/days within which  they must be used and the subsequent chained keys will be generated by  server and authenticator.  If the user does not use the key pairs within its validity period, the authenticator  will ask the user to re-login and will generate  new key pairs. The connection between the server and the authenticator  is also protected by implementing the quantum attack resistant public key algorithm CRYSTALS-KYBER~\cite{crystals} over TLS 1.3. (Cloudflare and Google have tested the feasibility in~\cite{cloudflareGooglePQC2020}). 
    Next, the user creates a request with the authenticator to add  a device. In the background, the server generates a transient token ($T_n$) using the Time based One-Time Password protocol (TOTP)~\cite{rfc6238}. It is the de-facto standard to generate a two factor authentication code. By default each TOTP expires after 30 seconds. The server sends the $(T_n)$ together with the API address ($S_a$) to the authenticator, encrypted with the authenticator's public key: $ S \rightarrow A :A^{S}_{p}(T_n,S_a)$.
    Upon decrypting  the token and the address, the authenticator  encrypts the token with server's  public key: $ S^{A}_{p}(T_{n})$ and signs it with its own private key: $ A^{S}_{s}(S^{A}_{p}(T_{n}))$. The authenticator sends the decrypted API address,  server's public key, and the signed-encrypted token to the device: $ A \rightarrow D: C_{K}(S_a,S^{A}_{p}, A^{S}_{s}(S^{A}_{p}(T_{n})))$. Note that this message is protected by $C_K $. 
    The device generates a key pair $ (D^{S}_{s}, D^{S}_{p}) $, appends its public  key with  pseudo UUID $(D_{u})$ and the signed-encrypted token. It encrypts this message with the server's public key: $ D \rightarrow S: S^{A}_{p}(D^{S}_{p} || D_{u} || A^{S}_{s}(S^{A}_{p}(T_{n})))   $. 
    When the server receives the above message, it decrypts and checks the authenticity of the message from the signature of the token. It also checks the  validity of the time based token. After proper validation, the server registers the device to its database.
     In reply to the API call, the server generates a new long-lived token $ (T_{D}) $, a fresh   key pair $ (S^{D}_{s}, S^{D}_{p}) $ specifically for the device and sends it encrypted with the device public key: $ S \rightarrow D: D^{S}_{p}(T_{D}, S^{D}_{p}) $. The server also notifies the authenticator that the device is now connected to the IoT analytics cloud platform: $ S \rightarrow A: A^{S}_{p}(D_{u}|| <connected>) $.

After the above steps, the device and the server have  dedicated key pairs $ (D^{S}_{s}, D^{S}_{p}), (S^{D}_{s}, S^{D}_{p}) $ to mutually authenticate and secure their future messages. The long-lived token $T_D$ is chained for subsequent messages and  serves to revoke the key if the user decides to decommission the device or resale. Fig.~\ref{fig:protocol_flow} summarises  a high level message flow between different entities of the protocol. In future work, the device offboarding protocol and the ASOP protocol's security analysis will be provided.

%% file: sections/5_conclusion.tex
\section{Conclusion}
\label {conclusion}

In this preliminary work, we present a sovereign and secure application-level device onboarding protocol (ASOP) for IoT devices. The protocol does not rely on manufacturers or any third-party certificate authority.  It is designed to make the device onboarding process smoother for consumer-level users who usually do not maintain public key pairs to manage electronic devices. The use of the newly standardised CRYSTALS  protocol suite ensures that the ASOP protocol is future  proof.